\documentclass{elsart}

\usepackage{amssymb}
\usepackage{amsmath}

\begin{document}

\begin{frontmatter}



\title{The integrals of motion for
the elliptic deformation of
the Virasoro and $W_N$ algebra}

\author{T.KOJIMA}
\address{
Department of Mathematics, College of Science and Technology,
\\
Nihon University,
Surugadai, 
Chiyoda-ku, 
Tokyo 101-0062, JAPAN}
\author{J.SHIRAISHI}
\address{
Graduate school of Mathematical Science, \\
University of Tokyo,
Komaba, Megro-ku, Tokyo 153-8914, JAPAN
}
\begin{abstract}
We review 
the free field realization
of the deformed Virasoro algebra $Vir_{q,t}$
and the deformed $W$ algebra $W_{q,t}(\widehat{gl_N})$.
We explicitly construct two classes of infinitly many
commutative operators 
${\cal I}_m$, ${\cal G}_m$, $(m \in {\mathbb N})$,
in terms of these algebras.
They can be regarded as the elliptic deformation of
the local and nonlocal integrals of motion for the
conformal field theory
\cite{BLZ,BLZ2,BLZ3,BHK,Kojima}.
This review is based on the works
\cite{FKSW, FKSW2, KS}.

\end{abstract}


\end{frontmatter}

\section{Introduction}
\label{}

The Korteweg-de Vries (KdV) equation
occupies a central place in the modern theory of completely 
integrable systems.
Because of its integrability,
the KdV equation has infinitly many conservation laws.
The Hamiltonian aspects of the KdV theory connected it to
the conformal field theory.
The quantization of the second Poisson bracket
$\{,\}_{P.B.}$ of the KdV gives rise to the Virasoro algebra :
$[L_m,L_n]=(m-n)L_{m+n}+\frac{1}{12}c_{CFT}m(m^2-1)\delta_{m+n,0}$.
The quantum field theory of the KdV theory becomes the conformal field theory
associated with the Virasoro algebra
\cite{BLZ, BLZ2, BLZ3}.
V.Bazhanov, S.Lukyanov, Al.Zamolodchikov
\cite{BLZ} constructed quantum field theoretical analogue
of the commuting transfer matrix ${\bf T}(z)$ acting the highest weight module
of the Virasoro algebra.
The commuting transfer matrix ${\bf T}(z)$ is constructed
as the trace of an image of the universal $R$-matrix associated with the quantum 
affine symmetry $U_q(\widehat{sl_2})$.
Hence the commutativity $[{\bf T}(z),{\bf T}(w)]=0$ is a direct consequence of
the Yang-Baxter relation.
We call
the coefficients of the asymptotic expansion of 
${\rm log}{\bf T}(z)$, $(z \to \infty)$,
the local integrals of motion for the Virasoro algebra.
They recover the conservation laws of the KdV 
in the classical limit $c_{CFT}\to \infty$.
We call the coefficients of the Taylor expansion of ${\bf T}(z)$
the nonlocal integrals of motion for the Virasoro algebra.
See also the generalization to the $W_N$ algebra \cite{BHK, Kojima}.

In this paper we construct the elliptic deformation of the integrals of motion
for the conformal field theory \cite{BLZ, BHK, Kojima}.
In this paper we  construct two classes of infinitly many
commutative operators 
${\cal I}_m$, ${\cal G}_m$, $(m \in {\mathbb N})$,
associated with the deformed Virasoro algebra and the deformed $W$-algebra
$W_{q,t}(\widehat{gl_N})$.
Because it is not so easy to calculate the trace of the image of the universal $R$-matrix
of the elliptic quantum group,
we prefer the completly differnt method of the construction for the integrals of motion
in the elliptic deformation of the conformal field theory.
Instead of considering the transfer matrix ${\bf T}(z)$,
we directly give the explicit formulae of
the integrals of motion ${\cal I}_n$ and ${\cal G}_n$
for the deformed $W$-algebra $W_{q,t}(\widehat{gl_N})$.
The commutativity of the intregrals of motion are not understood as direct consequence
of the Yang-Baxter relation.
They are understood as consequence of the commutative family of
the Feigin-Odesskii algebra \cite{FO}.

The organization of this paper is as follows.
In section 2 we give reviews on the deformed Virasoro algebra and the deformed $W$-algebra
$W_{q,t}(\widehat{gl_N})$.
In section 3 we give explicit formulae of
the integrals of motion for the deformed Virasoro algebra and the deformed $W$-algebra,
and state the main theorem.

\section{Elliptic deformation
of the Virasoro algebra and the $W_N$-algebra}
In this section we review 
the elliptic deformation
of the Virasoro algebra and the $W_N$-algebra.
We fix three parameters $x,r,s$ such that
$0<x<1$, ${\rm Re}(r)>0$ and ${\rm Re}(s)>0$.
Let us set $r^*=r-1$.
We set the parameters $\tau$ by
$x=\exp\left(-\pi \sqrt{-1}/r\tau\right)$
We relate two variables $z$ and $u$ by $z=x^{2u}$.
The symbol $[u]_r$ stands for the Jacobi theta function
\begin{eqnarray}
~[u]_r=x^{\frac{u^2}{r}-u}
\frac{\Theta_{x^{2r}}(z)}{(x^{2r};x^{2r})_\infty},
~~
\Theta_{q}(z)=(z;q)_\infty
(q/z;q)_\infty (q;q)_\infty,
\end{eqnarray}
where $(z;q)_\infty=\prod_{j=0}^\infty
(1-q^jz)$.
The elliptic theta function satisfies the quasi-periodicities,
\begin{eqnarray}
~~~[u+r]_r=-[u]_r,~~[u+r\tau]_r=-
e^{
-\pi\sqrt{-1}\tau
-2\pi \sqrt{-1}u/r}[u]_r.
\end{eqnarray}
The symbol $[a]$ stands for $q$-integer $[a]=
\frac{x^a-x^{-a}}{x-x^{-1}}$.
\subsection{Bosons}

For $N=2,3,4,\cdots$, we introduce the bosons $\beta_m^j,
(m\in {\mathbb Z}_{\neq 0};
1\leqq j \leqq N)$, which satisfy  the commuttion relation,
\begin{eqnarray}
~[\beta_n^i, \beta_m^j]=
n\frac{[(r-1)n]}{[rn]}
\delta_{n+m,0}
\times
\left\{
\begin{array}{cc}
\frac{[(s-1)n]}{[sn]} & (1\leqq i=j\leqq N)\\
-\frac{[n]}{[sn]}x^{s n~{\rm sgn(i-j)}} & (1\leqq i \neq j 
\leqq N)
\end{array}
\right.
\label{def:boson}
\end{eqnarray}
For $N=2,3,4,\cdots$, we introduce 
the zero-mode operators $P_\lambda$ and $Q_\lambda$.
Let $\epsilon_j$, $(1\leqq j \leqq N)$ be
an orthonormal basis in ${\mathbb R}^N$ relative to the standard
inner product $(~|~)$.
Let us set $\bar{\epsilon}_i=\epsilon_i-\epsilon,$
$\epsilon=\frac{1}{N}\sum_{j=1}^N \epsilon_j$.
Let us set $\alpha_j=\bar{\epsilon}_j-\bar{\epsilon}_{j+1}$.
Let $P_\lambda$, $Q_\lambda$ 
be the zero mode operators defined by
the commutation relation 
\begin{eqnarray}
~~~~~~~~~~~~~~~~~~~~~~[iP_\lambda,Q_\mu]=(\lambda|\mu),~~~~
(\lambda, \mu \in \sum_{j=1}^N {\mathbb Z}\bar{\epsilon}_j)
.\label{def:zero-mode}
\end{eqnarray}
The action of the Dynkin-diagram automorphism $\eta$ on
the bosons is given by
\begin{eqnarray}
&&\eta(\beta_m^1)=x^{-\frac{2s}{N}m}\beta_m^2,
\cdots,
\eta(\beta_m^{N-1})=x^{-\frac{2s}{N}m}\beta_m^N,
\eta(\beta_m^N)=x^{\frac{2s}{N}(N-1)m}\beta_m^1.
\end{eqnarray}
The action of the Dynkin-diagram automorphism $\eta$ on
the zero-mode operator is given by
\begin{eqnarray}
\eta(P_\lambda)=P_{\eta(\lambda)},~~
\eta(Q_\lambda)=Q_{\eta(\lambda)},~~
\eta(\bar{\epsilon}_j)=\bar{\epsilon}_{j+1},~~(1\leqq j \leqq N),
\end{eqnarray} 
where we understand $\bar{\epsilon}_1=\bar{\epsilon}_{N+1}$.
Let us introduce the Fock space ${\cal F}_{l,k}$,
$(l,k \in \sum_{j=1}^N
{\mathbb Z}\bar{\epsilon}_j)$, of the bosons,
generated by $\beta_{-m}^j, (m>0)$ over the vacuum vector
$|l,k\rangle$, $(l,k \in \sum_{j=1}^N
{\mathbb Z}\bar{\epsilon}_j)$,
\begin{eqnarray}
&&\beta_{-m}^j |l,k\rangle=0,~(m>0; j=1,2,\cdots,N),\\
&&P_\alpha
|l,k\rangle=
\left(\alpha\left|
l\sqrt{\frac{r}{r-1}}-k\sqrt{\frac{r-1}{r}}\right)|l,k
\right.\rangle,\\
&&|l,k\rangle
=\exp\left(\sqrt{\frac{r}{r-1}}Q_l-
\sqrt{\frac{r-1}{r}}Q_k\right)|0,0\rangle.
\end{eqnarray}
\subsection{Deformed $W$-algebra}

In this section,
we review the deformed Virasoro algebra and the deformed $W$ algebra 
$W_{q,t}(\widehat{gl_N})$, following \cite{SKAO, AKOS, O, FF, FKSW, FKSW2, KS}.

\begin{defn}~~
For $N=2,3,4,\cdots$,
the deformed $W$-algebra $W_{q,t}(\widehat{gl_N})$
is generated by the generators ${T}_m^{(j)}$,
$(1\leqq j \leqq N, m \in {\mathbb Z})$,
with the defining relations (\ref{def:dW}) of
the series $T_j(z)=\sum_{m \in {\mathbb Z}}T_m^{(j)}z^{-m}$.
\begin{eqnarray}
&&f_{i,j}(z_2/z_1)T_i(z_1)T_j(z_2)-
f_{j,i}(z_1/z_2)T_j(z_2)T_i(z_1)
\nonumber\\
&=&
c \sum_{k=1}^i
\prod_{l=1}^{k-1}\Delta(x^{2l+1})\left(
\delta\left(\frac{x^{j-i+2k}z_2}{z_1}\right)
f_{i-k,j+k}(x^{-j+i})T_{i-k}(x^{-k}z_1)T_{j+k}(x^kz_2)\right.
\nonumber\\
&-&\left.
\delta\left(\frac{x^{-j+i-2k}z_2}{z_1}\right)
f_{i-k,j+k}(x^{j-i})T_{i-k}(x^{k}z_1)T_{j+k}(x^{-k}z_2)
\right),~
(1\leqq i \leqq j \leqq N),\nonumber\\
\label{def:dW}
\end{eqnarray}
where we used the delta-function
$\delta(z)=\sum_{m \in {\mathbb Z}}z^m$.
Here we have set the constant $c$ and the structure functions
$\Delta(z)$ and $f_{i,j}(z)$, 
$(1\leqq i, j \leqq N)$ by
\begin{eqnarray}
&&c=-\frac{(1-x^{2r})(1-x^{-2(r-1)})}{(1-x^2)},~~
\Delta(z)=\frac{(1-x^{2r-1}z)(1-x^{1-2r}z)}{(1-xz)(1-x^{-1}z)},
\\
&&f_{i,j}(z)\\
&=&
\exp\left(
\sum_{m=1}^\infty
\frac{(1-x^{2rm})(1-x^{-2(r-1)m})
(1-x^{2m Min(i,j)})(1-x^{2m(s-Max(i,j))})
}{m (1-x^{2m})(1-x^{2sm})}(x^{|i-j|}z)^m
\right).\nonumber
\end{eqnarray}
\end{defn}
~\\
{\bf Example}~~Upon the specialization $N=s=2$,
we have the deformed Virasoro algebra $Vir_{q,t}$.
Upon this specialization the generators $T_m^{(2)}$
can be regarded as $T_m^{(2)}=1$.
The generators $T_m^{(1)}=T_m$ satisfy the following defining 
relation.
\begin{eqnarray}
~~~~~
\sum_{l=0}^\infty
f_l(T_{n-l}T_{m+l}-T_{m-l}T_{n+l})=c(c^{2n}-x^{-2n})
\delta_{n+m,0},
\end{eqnarray}
where the structure constant $f_l$ is given by
$\sum_{l=0}^\infty f_l z^l=f_{1,1}(z)$.
In the CFT limit $(x \to 1)$, we get the Virasoro algebra
with the central charge $c_{CFT}=1-\frac{6}{r(r-1)}$.
\begin{eqnarray}
~~~[L_m,L_n]=(m-n)L_{m+n}+\frac{1}{12}c_{CFT}~
m(m^2-1)\delta_{m+n,0}.
\end{eqnarray}

\begin{prop}~~
For $N=2$ the deformed $W$-algebra $W_{q,t}(\widehat{gl_2})$
is realized by the bosons (\ref{def:boson}),
(\ref{def:zero-mode}) on the Fock space.
\begin{eqnarray}
~~~T_1(z)=\Lambda_1(z)+\Lambda_2(z),~~
T_2(z)=:\Lambda_1(x^{-1}z)\Lambda_2(xz):,
\end{eqnarray}
where we have set
\begin{eqnarray}
\Lambda_1(z)&=&x^{-\sqrt{r (r-1)}P_{\alpha_1}}
:\exp\left(\sum_{m \neq 0}\frac{1}{m}(x^{rm}-x^{-rm})\beta_m^1
z^{-m}\right):,\\
\Lambda_2(z)&=&x^{-\sqrt{r (r-1)}P_{\alpha_2}}:\exp\left(
\sum_{m \neq 0}\frac{1}{m}(x^{rm}-x^{-rm})\beta_m^2
z^{-m}
\right):.
\end{eqnarray}
\end{prop}

\begin{prop}~~~
For $N=3,4,\cdots$ 
the deformed $W$-algebra $W_{q,t}(\widehat{gl_N})$
is realized by the bosons (\ref{def:boson}),
(\ref{def:zero-mode}) on the Fock space.
\begin{eqnarray}
T_j(z)=\sum_{1\leqq s_1<s_2<\cdots<s_j\leqq N}
:\Lambda_{s_1}(x^{-j+1}z)\Lambda_{s_2}(x^{-j+3}z)\cdots 
\Lambda_{s_j}(x^{j-1}z):,
\end{eqnarray}
where we have set
\begin{eqnarray}
\Lambda_j(z)=x^{-\sqrt{r(r-1)}P_{\bar{\epsilon}_j}}
:\exp\left(\sum_{m \neq 0}\frac{x^{rm}-x^{-rm}}{m}\beta_m^j
z^{-m}\right):, (1\leqq j \leqq N).
\end{eqnarray}
\end{prop}
Here the symbol $:*:$ stands for usual normal ordering of bosons,
i.e. $\beta_m^i$ with $m>0$ should be moved to the right.

\subsection{Screening current}

In this section we review the screening currents
for the deformed Virasoro algebra and the deformed $W$ algebra, following
\cite{SKAO, AKOS, O, FF, AJMP, FJMOP, KK, FKSW, FKSW2, KS}.

\begin{defn}~~
For $N=2$ we introduce 
the operator $F_j(z), (j=1,2)$, called the screening current
for the deformed $W$-algebra $W_{q,t}(\widehat{gl_2})$.
We define
\begin{eqnarray}
F_1(z)&=&e^{-i\sqrt{\frac{r^*}{r}}Q_{\alpha_1}}
z^{-\sqrt{\frac{r^*}{r}}P_{\alpha_1}+\frac{r^*}{r}}
:\exp\left(\sum_{m \neq 0}\frac{1}{m}
(\beta_m^1-\beta_m^2)z^{-m}\right):
,\\
F_2(z)&=&
e^{-i\sqrt{\frac{r^*}{r}}Q_{\alpha_2}}
z^{-\sqrt{\frac{r^*}{r}}P_{\alpha_2}+\frac{r^*}{r}}
:\exp\left(\sum_{m \neq 0}\frac{1}{m}
(-x^{sm}\beta_m^1+x^{-sm}\beta_m^2)z^{-m}\right):.
\nonumber\\
\end{eqnarray}
\end{defn}

\begin{defn}~~
For $N=3,4,\cdots$ we introduce 
the operator $F_j(z), (1\leqq j \leqq N)$, 
called the screening current
for the deformed $W$-algebra $W_{q,t}(\widehat{gl_N})$.
Let us set
\begin{eqnarray}
F_j(z)&=&e^{-i\sqrt{\frac{r}{r-1}}Q_{\alpha_j}}
(x^{(\frac{2s}{N}-1)j}z)^{-\sqrt{\frac{r}{r-1}}P_{\alpha_j}+
\frac{r}{r-1}}\nonumber\\
&\times&:\exp\left(
\sum_{m\neq 0}\frac{1}{m}
(\beta_m^j-\beta_m^{j+1})(x^{\frac{2s}{N}}z)^{-m}
\right):,~~(1\leqq j \leqq N-1),\\
F_N(z)&=&e^{-i\sqrt{\frac{r}{r-1}}Q_{\alpha_N}}
(x^{2s-N}z)^{-\sqrt{\frac{r}{r-1}}P_{\bar{\epsilon}_N}+
\frac{r}{2(r-1)}}
z^{\sqrt{\frac{r}{r-1}}P_{\bar{\epsilon}_1}+
\frac{r}{2(r-1)}}
\nonumber\\
&\times&:\exp\left(
\sum_{m\neq 0}\frac{1}{m}
(x^{-2sm}\beta_m^N-\beta_m^{1})z^{-m}
\right):.
\end{eqnarray}
\end{defn}

\begin{prop}~~For $N=2$ 
the screening currents $F_1(z), F_2(z)$ satisfy
the following commutation relations.
\begin{eqnarray}
&&[u_1-u_2]_r[u_1-u_2+1]_rF_j(z_1)F_j(z_2)\nonumber\\
&&=
[u_2-u_1]_r[u_2-u_1+1]_rF_j(z_2)F_j(z_1),
~~~~(j=1,2),\\
&&
\left[u_1-u_2+\frac{s}{2}-1\right]_r
\left[u_1-u_2-\frac{s}{2}\right]_r
F_2(z_1)F_1(z_2)\nonumber\\
&&
=\left[u_2-u_1+\frac{s}{2}-1\right]_r
\left[u_2-u_1-\frac{s}{2}\right]_r
F_1(z_2)F_2(z_1).
\end{eqnarray}
\end{prop}

\begin{prop}~~
For $N=2$ the commutation relations
between $\Lambda_j(z)$ and $F_j(z)$ are given by
\begin{eqnarray}
&&[\Lambda_1(z_1),F_1(z_2)]=(x^{-r^*}-x^{r^*})
\delta(x^rz_1/z_2){\cal A}(x^{-r}z_2),\\
&&[\Lambda_2(z_1),F_1(z_2)]=(x^{r^*}-x^{-r^*})
\delta(x^{-r}z_1/z_2){\cal A}(x^{r}z_2),\\
&&[\Lambda_1(z_1),F_2(z_2)]=(x^{r^*}-x^{-r^*})
\delta(x^{-r+s}z_1/z_2)\eta({\cal A}(x^{r}z_2)),\\
&&[\Lambda_2(z_1),F_2(z_2)]=
(x^{-r^*}-x^{r^*})\delta(x^{r-s}z_1/z_2)\eta({\cal A}(x^{-r}z_2)),
\end{eqnarray}
where we have set
\begin{eqnarray}
{\cal A}(z)=e^{i\sqrt{\frac{r^*}{r}}Q_{\alpha_1}}
z^{\sqrt{\frac{r^*}{r}}P_{\alpha_1}+\frac{r^*}{r}}
:\exp\left(\sum_{m \neq 0}\frac{1}{m}
(x^{rm}\beta_m^1-x^{-rm}\beta_m^2)z^{-m}
\right):.\nonumber\\
\end{eqnarray}
\end{prop}

\begin{prop}~~For $N=3,4,5,\cdots$,
the screening currents $F_j(z)$
satisfy the following commutation relations.
\begin{eqnarray}
&&\left[u_1-u_2-\frac{s}{N}\right]_r
F_j(z_1)F_{j+1}(z_2)=
\left[u_2-u_1+\frac{s}{N}-1\right]_r
F_{j+1}(z_2)F_{j}(z_1),\nonumber\\
\\
&&\left[u_1-u_2\right]_r
\left[u_1-u_2+1\right]_r
F_j(z_1)F_j(z_2)=
\left[u_2-u_1\right]_r
\left[u_2-u_1+1\right]_r
F_j(z_2)F_j(z_1),\nonumber\\
\end{eqnarray}
for $1\leqq j \leqq N$. We understand $F_{N+1}(z)=F_1(z)$.
We have
\begin{eqnarray}
F_i(z_1)F_j(z_2)=F_j(z_2)F_i(z_1),~~
{\rm otherwise}.
\end{eqnarray}
\end{prop}

\begin{prop}~~
For $N=3,4,5,\cdots$ the commutation relations
between $\Lambda_j(z)$ and $F_j(z)$ are given by
\begin{eqnarray}
&&[\Lambda_j(z_1),F_j(z_2)]=(x^{-r^*}-x^{r^*})
\delta(x^{-\frac{2s}{N}j+r}z_1/z_2)
{\cal A}_j(x^{\frac{2s}{N}j-r}z_1),~(1\leqq j \leqq N-1),
\nonumber\\
\\
&&[\Lambda_{j+1}(z_1),F_j(z_2)]=(x^{r^*}-x^{-r^*})
\delta(x^{-\frac{2s}{N}j-r}z_1/z_2){\cal A}_j
(x^{\frac{2s}{N}j+r}z_2),~(1\leqq j \leqq N-1),\nonumber\\
\\
&&[\Lambda_N(z_1),F_N(z_2)]=(x^{-r^*}-x^{r^*})
\delta(x^{r-2s}z_1/z_2){\cal A}_N(x^{-r}z_2),\\
&&[\Lambda_1(z_1),F_N(z_2)]=
(x^{r^*}-x^{-r^*})\delta(x^{-r}z_1/z_2){\cal A}_N(x^{r}z_2),
\end{eqnarray}
where we have set
\begin{eqnarray}
{\cal A}_j(z)&=&
e^{i\sqrt{\frac{r^*}{r}}Q_{\alpha_j}}
x^{-\sqrt{r r^*}
(P_{\bar{\epsilon}_j}+
P_{\bar{\epsilon}_{j+1}})}
(x^{-j}z)^{\sqrt{\frac{r^*}{r}}P_{\alpha_j}+
\frac{r^*}{r}}
\nonumber\\
&\times&
:\exp\left(\sum_{m \neq 0}\frac{1}{m}
(x^{rm}\beta_m^j-x^{-rm}\beta_m^{j+1})z^{-m}
\right):,~(1\leqq j \leqq N-1),\\
{\cal A}_N(z)&=&
e^{i\sqrt{\frac{r^*}{r}}Q_{\alpha_N}}
x^{-\sqrt{r r^*}
(P_{\bar{\epsilon}_N}+
P_{\bar{\epsilon}_{1}})}
(x^{2s-N}z)^{\sqrt{\frac{r^*}{r}}P_{{\bar{\epsilon}_N}}+
\frac{r^*}{2r}}
z^{-\sqrt{\frac{r^*}{r}}P_{{\bar{\epsilon}_1}}+
\frac{r^*}{2r}}
\nonumber\\
&\times&
:\exp\left(\sum_{m \neq 0}\frac{1}{m}
(x^{(r-2s)m}\beta_m^N-x^{-rm}\beta_m^{1})z^{-m}
\right):.
\end{eqnarray}
\end{prop}

\section{Integrals of Motion}
\label{}

In this section we review the integrals of motion
for the deformed Virasoro algebra and
the deformed $W$-algebra, following
\cite{FKSW, FKSW2, KS}.

\subsection{Local integrals of motion ${\cal I}_n$}

We define the operators ${\cal I}_n$,
$(n=1,2,3,\cdots)$,
which we call the local integrals of motion for 
the $W_{q,t}(\widehat{gl_N})$, $(N=2,3,4,\cdots)$.

\begin{defn}~~
For the regime ${\rm Re}(s)>2$ and ${\rm Re}(r^*)<0$,
we define
\begin{eqnarray}
{\cal I}_n&=&\int \cdots \int_C \prod_{j=1}^n
\frac{dz_j}{z_j}T_1(z_1)T_1(z_2)T_1(z_3)
\cdots T_1(z_n)\nonumber\\
&\times&
\prod_{1\leqq j<k \leqq n}\frac{[u_k-u_j]_s
[u_k-u_j+r]_s}{[u_k-u_j+1]_s[u_k-u_j+r^*]_s}, 
~~~(n=1,2,3,\cdots).
\end{eqnarray}
Here the contour $C$ encircles $z_j=0$
in such a way that $z_j=x^{-2+2sl}z_k,
x^{-2r^*+2sl}z_k$, $(l=0,1,2,\cdots)$
is inside and
$z_j=x^{2-2sl}z_k,
x^{2r^*-2sl}z_k$, $(l \in {\bf N})$
is outside for $1\leqq j<k \leqq n$.
We call ${\cal I}_n$ 
the local integrals of motion.
\end{defn}
The definitions of the local integrals of motion
${\cal I}_n$
for generic ${\rm Re}(s)>0$ and ${\rm Re}(r)>0$
should be understood as analytic continuation.

\subsection{Nonlocal integrals of motion ${\cal G}_n$}

We define the operators ${\cal G}_m$,
$(m=1,2,3,\cdots)$,
which we call the nonlocal integrals of motion for 
the $W_{q,t}(\widehat{gl_N})$, $(N=2,3,4,\cdots)$.

\begin{defn}~~
For $N=2$ and the regime $0<{\rm Re}(s)<2$ and
${\rm Re}(r)>0$, we define
\begin{eqnarray}
{\cal G}_m
&=&\int \cdots \int_{I}
\prod_{t=1,2} \prod_{j=1}^m
\frac{dz_j^{(t)}}{z_j^{(t)}}F_1(z_1^{(1)})F_1(z_2^{(1)})
\cdots F_1(z_m^{(1)})
F_2(z_1^{(2)})F_2(z_2^{(2)})\cdots F_2(z_m^{(2)})
\nonumber\\
&\times&
\frac{
\displaystyle
\prod_{t=1,2}
\prod_{1\leqq j<k \leqq m}
\left[u_j^{(t)}-u_k^{(t)}\right]_r
\left[u_k^{(t)}-u_j^{(t)}-1\right]_r
}{
\displaystyle
\prod_{j,k=1}^m
\left[u_j^{(1)}-u_k^{(2)}+\frac{s}{2}\right]_r
\left[u_k^{(2)}-u_j^{(1)}+\frac{s}{2}-1\right]_r
}\\
&\times&\prod_{t=1,2}
\left[\sum_{j=1}^m(u_j^{(t)}-u_j^{(t+1)})-\sqrt{r r^*}P_{\bar{\epsilon}_{t+1}}
\right]_r,~(m=1,2,\cdots).\nonumber
\end{eqnarray}
Here the contour $I$ 
encircles $z_j^{(t)}=0$, $(t=1,2; 1\leqq j \leqq m)$
in such a way that 
\begin{eqnarray}
|x^{s+2lr}z_k^{(2)}|, |x^{2-s+2lr}z_k^{(2)}|
<|z_j^{(1)}|<
|x^{-s-2lr}z_k^{(2)}|, |x^{s-2-2lr}z_k^{(2)}|,
\end{eqnarray}
for $1\leqq j, k \leqq m$ and $l \in {\mathbb N}$.
We call ${\cal G}_m$ 
the nonlocal integrals of motion for the deformed $W$-algebra 
$W_{q,t}(\widehat{gl_2})$.
\end{defn}

\begin{defn}~~
For $N=3,4,5,\cdots$ and the regime $0<{\rm Re}(s)<N$ and
${\rm Re}(r)>0$, we define
\begin{eqnarray}
{\cal G}_m
&=&\int \cdots \int_{I}
\prod_{t=1}^N \prod_{j=1}^m
\frac{dz_j^{(t)}}{z_j^{(t)}}F_1(z_1^{(1)})F_1(z_2^{(1)})
\cdots F_1(z_m^{(1)})\nonumber\\
&\times&
F_2(z_1^{(2)})F_2(z_2^{(2)})\cdots F_2(z_m^{(2)})
\cdots
F_N(z_1^{(N)})F_N(z_2^{(N)})\cdots F_N(z_m^{(N)})
\nonumber\\
&\times&
\frac{
\displaystyle
\prod_{t=1}^N
\prod_{1\leqq j<k \leqq m}
\left[u_j^{(t)}-u_k^{(t)}\right]_r
\left[u_k^{(t)}-u_j^{(t)}-1\right]_r
}{
\displaystyle
\prod_{t=1}^{N-1}
\prod_{j,k=1}^m
\left[u_j^{(t)}-u_k^{(t+1)}+1-\frac{s}{N}\right]_r
\prod_{j,k=1}^m
\left[u_j^{(1)}-u_j^{(N)}+\frac{s}{N}\right]_r
}\\
&\times&\prod_{t=1}^N
\left[\sum_{j=1}^m(u_j^{(t)}-u_j^{(t+1)})-\sqrt{r r^*}P_{\bar{\epsilon}_{t+1}}
\right]_r,~(m=1,2,\cdots).\nonumber
\end{eqnarray}
Here the contour $I$
encircles $z_j^{(t)}=0$, $(1\leqq t\leqq N; 1\leqq j \leqq m)$
in such a way that 
\begin{eqnarray}
&&|x^{\frac{2s}{N}+2lr}z_k^{(t+1)}|
<|z_j^{(t)}|<
|x^{-2+\frac{2s}{N}-2lr}z_k^{(t+1)}|,~(1\leqq t \leqq N-1),\\
&&|x^{2-\frac{2s}{N}+2lr}z_k^{(1)}|
<|z_j^{(N)}|<
|x^{-\frac{2s}{N}-2lr}z_k^{(1)}|,
\end{eqnarray}
for $1\leqq j, k \leqq m$ and $l \in {\mathbb N}$.
We call ${\cal G}_m$ 
the nonlocal integrals of motion for the deformed $W$-algebra 
$W_{q,t}(\widehat{gl_N})$.
\end{defn}

The definitions of the nonlocal integrals of motion
${\cal G}_n$
for generic ${\rm Re}(s)>0$ and ${\rm Re}(r)>0$
should be understood as analytic continuation.

\subsection{Main results}

In this section we state the main results.

\begin{thm}~~For $N=2,3,4,\cdots$, 
the local integrals of motion
${\cal I}_n$ and the nonlocal integrals of motion
${\cal G}_m$ commute with each other. 
\begin{eqnarray}
~~~~[{\cal I}_n,{\cal I}_m]=[{\cal I}_n,{\cal G}_m]
=[{\cal G}_n,{\cal G}_m]=0,~~~(m,n=1,2,\cdots).
\end{eqnarray}
\end{thm}

These commutativities are understood as consequence of 
commuting family of the Feigin-Odesskii algebra
\cite{FO}.

\begin{thm}~~For $N=2,3,4,\cdots$, 
the local integrals of motion ${\cal I}_n$ and
the nonlocal integrals of motion
${\cal G}_n$ are invariant under the action of
the Dynkin-diagram automorphism $\eta$.
\begin{eqnarray}
~~~~~~~\eta({\cal I}_n)={\cal I}_n,~~~\eta({\cal G}_n)={\cal G}_n,~~~(n=1,2,\cdots).
\end{eqnarray}
\end{thm}

\section*
{\bf Acknowledgement}
We would like to
thank Prof.B.Feigin, 
Prof.M.Jimbo and 
Mr.H.Watanabe for useful 
communications.
We would like to thank 
Professors V.Bazhanov,
A.Belavin,
P.Bouwknegt, 
A.Chervov,
S.Duzhin, 
V.Gerdjikov,
K.Hasegawa,
P.Kulish,
W-X.Ma,
V.Mangazeev,~
K.Takemura~ 
and
M.Wadati 
for their interests in this work.
This work is partly supported by
Grant-in Aid for Young Scientist {\bf B}(18740092) and
Grant-in Aid for Scientific Research {\bf C}(16540183)
from JSPS.




\end{document}